\documentclass[twocolumn,showpacs,fleqn,nobibnotes]{revtex4}

\usepackage{amsmath}
\usepackage{graphicx}
\usepackage{float}
\usepackage{subfigure}
\usepackage{epsfig}
\newcommand{\rr}{\mbox{\boldmath $r$}}
\newcommand{\rrn}{\mbox{$r$}}

\newcommand{\gsim}{\raisebox{-0.5mm}{$\stackrel{>}{\scriptstyle{\sim}}$}}

\begin{document}

\title{QCD evolution and skewedness effects in color dipole description of DVCS}
\author{L. Favart
$^{1,a}$\footnotetext{$^1$E-mail:lfavart@ulb.ac.be}
and M.V.T. Machado $^{2,b,c,d}$\footnotetext{$^2$E-mail:magnus@if.ufrgs.br, magnus@ufpel.edu.br} }

\affiliation{$^a$ IIHE - CP 230, Universit\'e Libre de Bruxelles. 
  1050 Brussels, Belgium\\ $^b$ Instituto de F\'{\i}sica e Matem\'atica, Universidade Federal de
Pelotas\\
Caixa Postal 354, CEP 96010-090, Pelotas, RS, Brazil\\
$^c$ High Energy Physics Phenomenology Group, GFPAE,  IF-UFRGS \\
Caixa Postal 15051, CEP 91501-970, Porto Alegre, RS, Brazil\\
$^d$ CERN Theory Division. CH-1211 Gen\`eve 23, Switzerland
}

\begin{abstract}
 We show the role played by  QCD evolution and skewedness effects in the DVCS 
 cross section at large $Q^2$ within the color dipole description of the 
 process at photon level. The dipole cross section is given by the 
 saturation model, which can be  improved by DGLAP evolution at high 
 photon virtualities. We investigate both possibilities as well as the 
 off-forward effect through a simple phenomenological parametrisation. 
 The results are compared to the recent ZEUS DVCS data.
\end{abstract}

\maketitle

\section{Introduction}

 
An important clean process allowing us to access off-diagonal (skewed)
parton distributions, which carry new information on the nucleon's
dynamical degrees of freedom, is the Deeply Virtual Compton Scattering
(DVCS) \cite{Adloff:2001cn,Favart:2003kw,Chekanov:2003ya}. This is due
to the real photon in the final state being  an elementary (point-like)
particle rather than a bound state like a meson or more complicated
configurations. The skewed parton distributions are generally defined
via the Fourier transform of matrix elements of renormalized,
non-local twist-two operators (for a pedagogical view, see
Refs.~\cite{Freund:2002ff,Diehl:2003ny}). These composite operators
contain only two elementary fields of the theory, which are placed at
different positions becoming then non-local and operating in unequal
momentum nucleon states.

Hence skewedness takes into account dynamical correlations between
partons with different momenta. The high energy situation at HERA gives
the important opportunity to constrain them as well as to study the
evolution with virtuality of the resulting quark and gluon distributions.
There are several representations for skewed parton distributions
\cite{SKREP1,SKREP2,SKREP3,SKREP4,SKREP5}, which can be used to compute
the relevant observables in DVCS (or other exclusive processes) through a
factorization theorem \cite{dvcsfact}. They are input in numerical
solutions of the renormalization group or evolution equations (see e.g.
\cite{RGEs}), producing very reliable predictions up to NLO level
\cite{FreundMcDermott}

 On the other hand, the color dipole models have also been successful in
describing DVCS observables
\cite{Donnachie:2000px,McDermott:2001pt,Favart:2003cu}. There, the main
degrees of freedom are the color dipoles, which interact with the nucleon
target via gluonic exchange. This interaction is modeled through the
dipole-nucleon cross section, which can include QCD dynamical effects
given by DGLAP, BFKL or non-linear high energy evolution equations (parton
saturation). Skewedness effects are not considered in the current dipole
models and this is one of the goals of the present analysis, making use of
a simple phenomenological parametrisation to estimate them. Moreover, the
QCD DGLAP evolution can be introduced, which improves the data description
in the large $Q^2$ kinematic region accessible in the recent ZEUS DVCS
measurements \cite{Chekanov:2003ya}.


This note is organized as follows. In the next section, we recall the main
formulas for the color dipole formalism applied to DVCS. For the dipole
cross section we have considered the saturation model
\cite{Golec-Biernat:1998js}, which produces a unified and intuitive
description of DIS \cite{Golec-Biernat:1998js}, diffractive DIS
\cite{Golec-Biernat:1999qd}, vector meson production
\cite{Caldwell:2001ky}, Drell-Yan \cite{Betemps:2001he,Betemps:2003je} and
DVCS \cite{Favart:2003cu}. In particular, the restriction to the
transverse part of the photon wave function, due to the real final state
photon in DVCS, enhances the contribution of larger dipole configurations
and therefore the sensitivity to soft content and to the transition
between hard/soft regimes. Such a feature provides a particularly relevant
test of saturation models. Moreover, the approach includes all twist
resummation, in contrast with the leading twist approximations. In the
Sec. 3, we discuss the role played by the QCD evolution and skewedness in
the high virtuality kinematic region. We also perform a systematic
analysis in order to investigate to what extent the distinct models
improve the data description. These issues have implications in the
correct determination of the $t$ slope parameter $B$, whose value has
never been measured for DVCS. Finally, the last section summarizes our
main results.

\section{DVCS cross section in dipole picture} 


In the proton rest frame, the DVCS process can be seen as 
a succession in time of three factorisable subprocesses: i) 
the photon fluctuates in a quark-antiquark pair, ii) this 
color dipole interacts with the proton target, iii) the quark pair
annihilates in a real photon. The usual kinematic variables are the
$\gamma^* p$ c.m.s. energy squared $s=W^2=(p+q)^2$,
where $p$ and $q$ are the proton and the photon
momenta respectively, the photon virtuality squared $Q^2=-q^2$ and
the Bjorken scale $x_{\mathrm{Bj}}=Q^2/(W^2+Q^2)$. 

The imaginary part of the DVCS amplitude at zero momentum transfer in the color dipole formalism  is expressed in the simple way \cite{Favart:2003cu}, 
\begin{eqnarray} 
& &   {\cal I}m\, {\cal A}\,(s,t=0)  =  \int \limits_0^1 dz \int\limits_{0}^ {\infty} d^2\rr\, 
H(z,\rr,Q^2)\,\sigma_{dip}(\tilde{x},\rr^2)\label{dvcsdip}\\
& &   H  =  \frac{6\alpha_{\mathrm{em}}}{4\,\pi^2}  \sum_f e_f^2 \, \left\{[z^2 +
(1-z)^2]\, \varepsilon_1 \,K_1 (\varepsilon_1 \,\rrn) \,\varepsilon_2
\,K_1 (\varepsilon_2 \,\rrn) \right. \nonumber \\
& & \,\,\,\,\,\, \, + \,\, \,\,\left. m_f^2 \, \,K_0(\varepsilon_1\,
\rrn)\,K_0(\varepsilon_2\, \rrn)  \right\}\,,\label{wdvcstrans}
\end{eqnarray} 
where $H(z,\rr,Q_{1,2}^2)=\Psi_T^*(z,\,\rr,\,Q_1^2=Q^2)\,
\Psi_T(z,\,\rr,\,Q_2^2=0)$, with $\Psi_T$  being the light cone  photon wave
function for transverse photons . Here, $Q_1=Q$ is the virtuality of the
incoming photon, whereas $Q_2$ is the virtuality of the outgoing real
photon. The longitudinal piece does not contribute at $Q_2^2=0$. The
relative transverse separation of the pair (dipole) is labeled by $\rr$
and $z$,
$(1-z)$, are the longitudinal momentum fractions of the quark (antiquark).
The auxiliary variables $\varepsilon^2_{1,\,2}= z(1-z)\,Q_{1,\,2}^2 +
m_f^2$ depend on the quark mass, $m_f$. The $K_{0,1}$ are the McDonald
functions and summation is taken over the quark flavors.


Let us summarize the main features and expressions from the saturation
model, which will be used here to estimate the DVCS cross section.  A
previous analysis compared to H1 data can be found in 
Ref.~\cite{Favart:2003cu}. The saturation model reproduces color transparency
behavior, $\sigma_{dip}\sim \rr^2$, for small dipoles, whereas it gives a
constant behavior for large ones. This is rendered by a dipole cross
section having an eikonal-like form,
\begin{eqnarray} 
 &\sigma_{dip} (\tilde{x}, \,\rr^2) =  \sigma_0 \,
 \left[\, 1- \exp \left(-\frac{\,Q_{\mathrm{sat}}^2(\tilde{x})\,\rr^2}{4} 
  \right) \, \right]\,,
 \label{gbwdip}\\ 
  &Q_{\mathrm{sat}}^2(\tilde{x}) = \left(\frac{x_0}{\tilde{x}}
  \right)^{\lambda} \,\mathrm{GeV}^2,\,\,\, \tilde{x}= x_{\mathrm{Bj}} 
  \!\left( \, 1+ \frac{4\,m_f^2}{Q^2} \,\right),
\end{eqnarray} 
where the saturation scale $Q_{\mathrm{sat}}(x)$ (energy dependent)  
defines the onset of the saturation phenomenon and sets the interface
between soft/hard domains. The parameters were obtained from a fit to the
HERA data producing $\sigma_0=23.03 \,(29.12)$ mb, $\lambda= 0.288 \,
(0.277)$ and $x_0=3.04 \cdot 10^{-4} \, (0.41 \cdot 10^{-4})$ for a
3-flavor (4-flavor) analysis~\cite{Golec-Biernat:1998js}. 
An additional parameter is the effective
light quark mass, $m_f=0.14$ GeV. For the 4-flavor analysis, the charm
quark mass is considered to be $m_c=1.5$ GeV.


 The QCD evolution to the original saturation model was implemented
recently \cite{Bartels:2002cj} (BGBK), where the dipole cross section now
depends on the gluon distribution in a Glauber-Gribov inspired way,
\begin{eqnarray}
 \sigma_{dip}\, (\tilde{x}, \rr^2)  =  \sigma_0 \! \left[ 1- \exp
\left(-\frac{\,\pi^2\,\rr^2\,\alpha_s(\mu^2)\,\tilde{x}\,G(\tilde{x},\mu^2)}{3\,\sigma_0} \right)  \right],
\label{bgkdip} 
\end{eqnarray} 
where the energy scale is defined as $\mu^2=C/\rr^2 + \mu_0^2$. The
parameters are determined from a fit to DIS data, with the following
initial condition for LO DGLAP evolution, $x\,G(x,\mu^2=1\,
\mbox{GeV}^2)=A_g\,x^{-\lambda_g}\,(1-x)^{5.6}$. The flavor number is
taken to be equal to 3.
The overall normalization $\sigma_0=23.03$ mb is kept fixed (labeled fit 1 in
Ref.~\cite{Bartels:2002cj}).  The DGLAP evolution improves the data
description in large $Q^2$ regime and brings the model close to the
theoretical high energy non-linear QCD approaches.

\begin{figure}[t]
\psfig{file=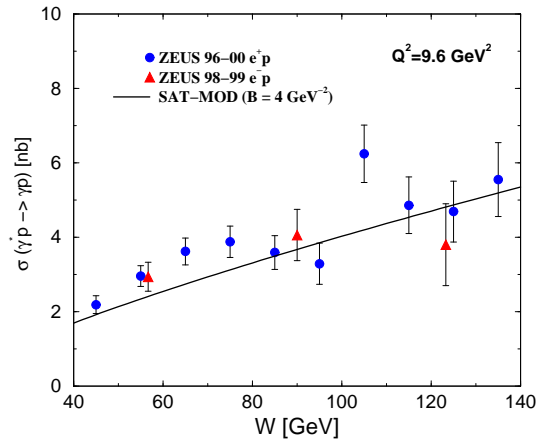,width=70mm} 
 \caption{The DVCS cross section as a function of c.m.s. energy, 
  $W_{\gamma p}$. The curve is the result for the saturation model for 
  fixed slope $B=4$ GeV$^{-2}$.}
\label{fig:1} 
\end{figure} 

Having a suitable model for the dipole cross section, as in Eq.
(\ref{gbwdip}) or Eq. (\ref{bgkdip}), we can use Eq. (\ref{dvcsdip}) and
then compute the final expression for the DVCS cross section as,
\begin{eqnarray} 
 \sigma(\gamma^*\,p\rightarrow \gamma \,p) & = & 
  \frac{1}{B}\,\frac{[\,{\cal I}m\,{\cal A}(s,0)\,]^2}{16\,\pi} \, 
    \left(1+\rho^2 \right) \label{dvcssigma}\,,
\end{eqnarray} 
where $B$ is the $t$ slope parameter (the behavior in $|t|$ is supposed
to obey a simple exponential parametrisation). 

In our further calculations, the real part is included via the usual
estimate $\rho=\tan (\pi\lambda /2)$, where $\lambda=\lambda (Q^2)$ is the
effective power of the imaginary part of the amplitude. We have fitted it for
$1\leq Q^2\leq 100$ GeV$^2$ in the form $\lambda_{\mathrm{eff}}(Q^2)=0.2 +
0.0107\ln^2(Q^2/2.48)$. It can be verified that when it rises to $\sim
0.3$ at high virtualities the contribution of the real part can reach
20\% of the total cross section.  In the next section we compute the cross
section above using the two versions for the saturation model and contrast
them with the recent DVCS ZEUS data, which includes data points with
larger $Q^2$ values than the previous H1 data. Moreover, we present a
simple way to introduce skewedness effects into the calculation.

\section{ Results and discussions} 

\begin{figure*}[t]
\begin{tabular}{cc}
\psfig{file=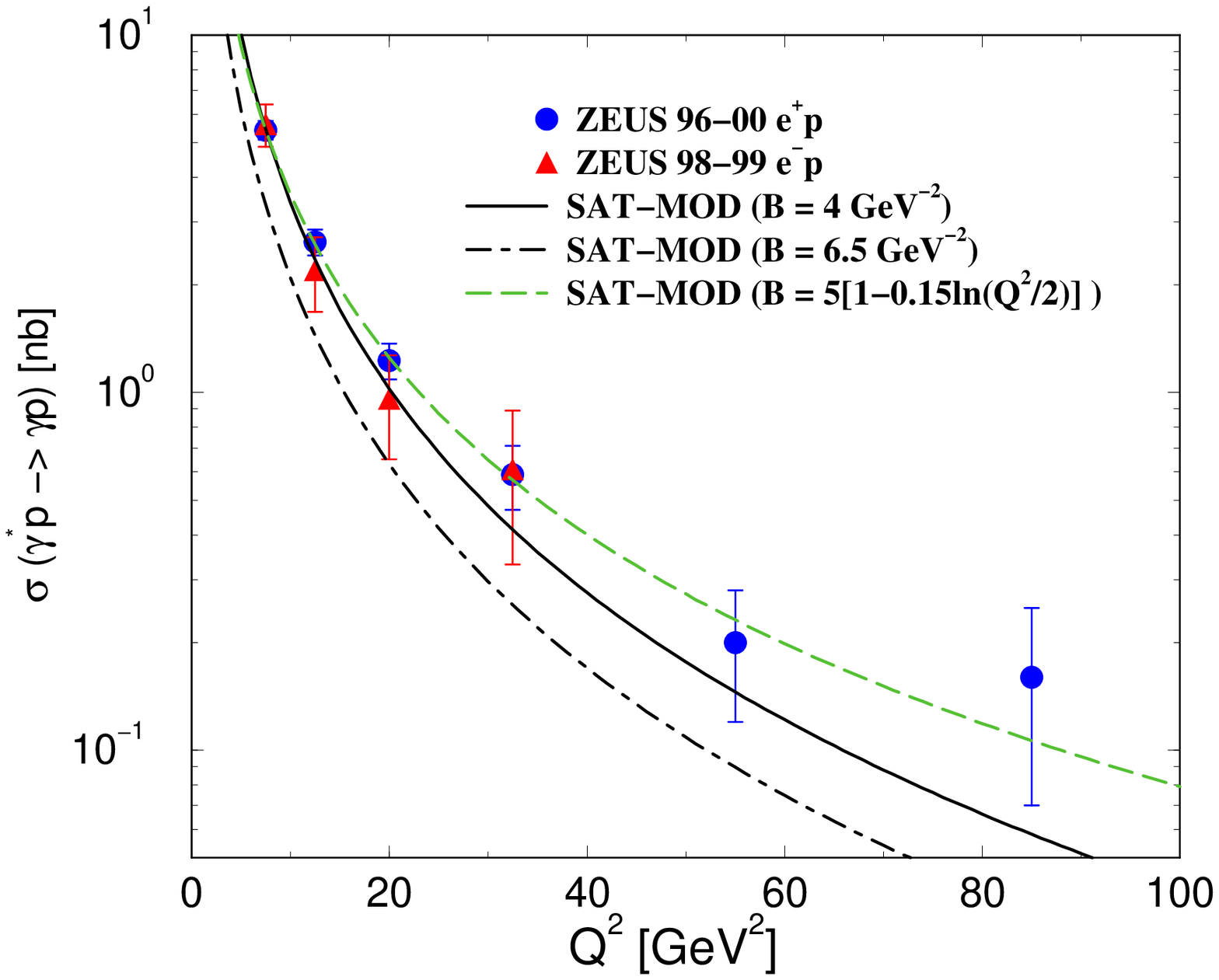,width=75mm} & \psfig{file=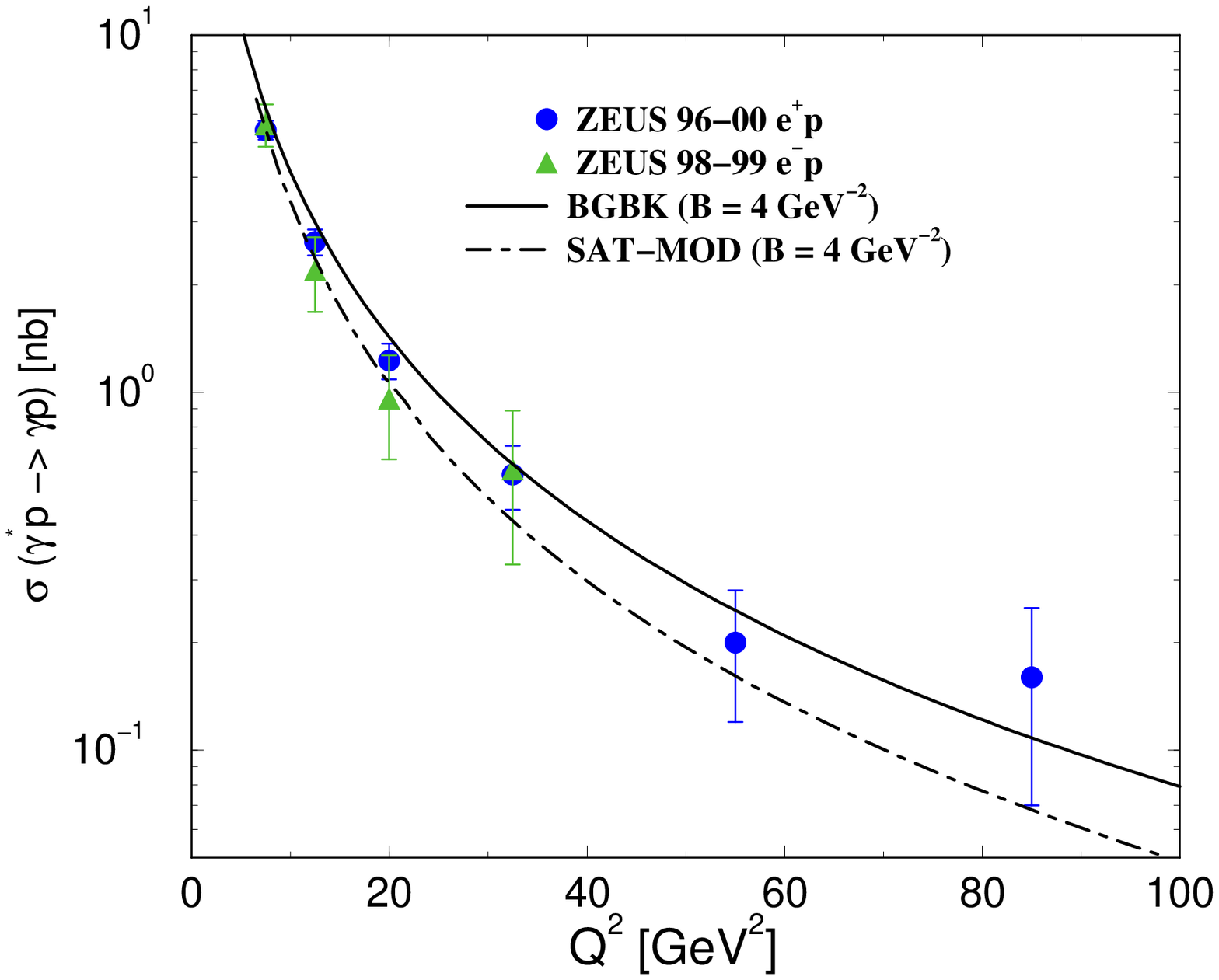,width=75mm}\\
(a) & (b)
\end{tabular}
 \caption{The DVCS cross section as a function of photon virtuality: (a) 
  saturation model using $B=4$ and 6.5 GeV$^{-2}$ (solid and dashed 
  curves) and $Q^2$-dependent slope (dot-dashed curve - see text). 
  (b) Effect of the BGBK model (includes 
  QCD evolution) using $B = 4$ GeV$^{-2}$ (solid and dot-dashed 
  curves).}
\label{fig:2} 
\end{figure*} 

In Fig.~\ref{fig:1} is shown the result for the saturation model, Eq.
(\ref{gbwdip}), confronted to the experimental data on DVCS of recent ZEUS
measurements as a function of the c.m.s. energy, $W_{\gamma p}$ (at fixed
virtuality $Q^2=9.6$ GeV$^2$). The parameters of the 4-flavor fit have
been used, producing good agreement with a fixed value for the slope,
$B=4$ GeV$^{-2}$.

In Fig.~\ref{fig:2}-a, we show the result of the saturation model for the
behavior with $Q^2$ at fixed energy, $W_{\gamma p}=89$ GeV.
In order to illustrate the sensitivity on the slope value, both values 
$B=4$ GeV$^{-2}$ (solid line) and $B=6.5$ GeV$^{-2}$ (dot-dashed line) are
shown$^1$\footnotetext[1]{It is worth mentioning that a slope
$B=6.5$ GeV$^{-2}$ (4-flavor) was able to describe correctly the H1 
experimental data for $Q^2\leq 40$ GeV$^2$ \cite{Favart:2003cu}.}. 
Although the statistical errors are large, it seems that for $Q^2\gsim 40 $ 
GeV$^2$, the model underestimates the experimental data. 
This can indicate two things: 
(a) the slope diminishes as the virtuality increases or, 
(b) some additional effect appears at higher $Q^2$.
In order to investigate the first hypothesis, we compute the cross section
using a $Q^2$ dependent slope, proposed in Ref.~\cite{Freund:2002qf}. That
is, $B(Q^2)= B_0\,[1-0.15\ln(Q^2/2)]$ GeV$^{-2}$ which is based on the 
diffractive electroproduction of $\rho$. 
Such a slope dependence allows a good description of the  $Q^2$
dependence of the cross section up to the highest measured values 
and gives a good normalisation for $B_0=5$ GeV$^{-2}$.

\begin{figure*}[t]
\begin{tabular}{cc}
\psfig{file=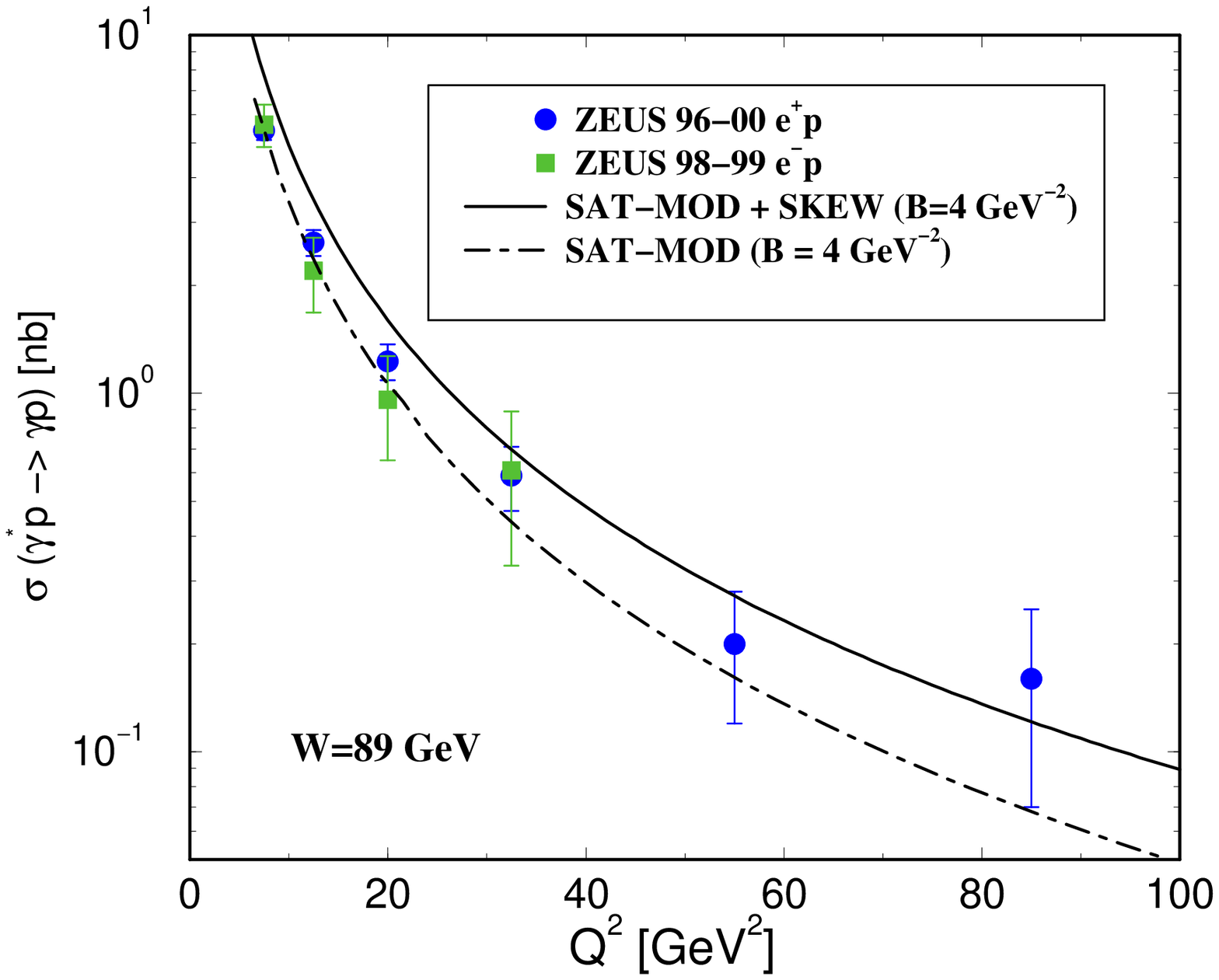,width=75mm} & \psfig{file=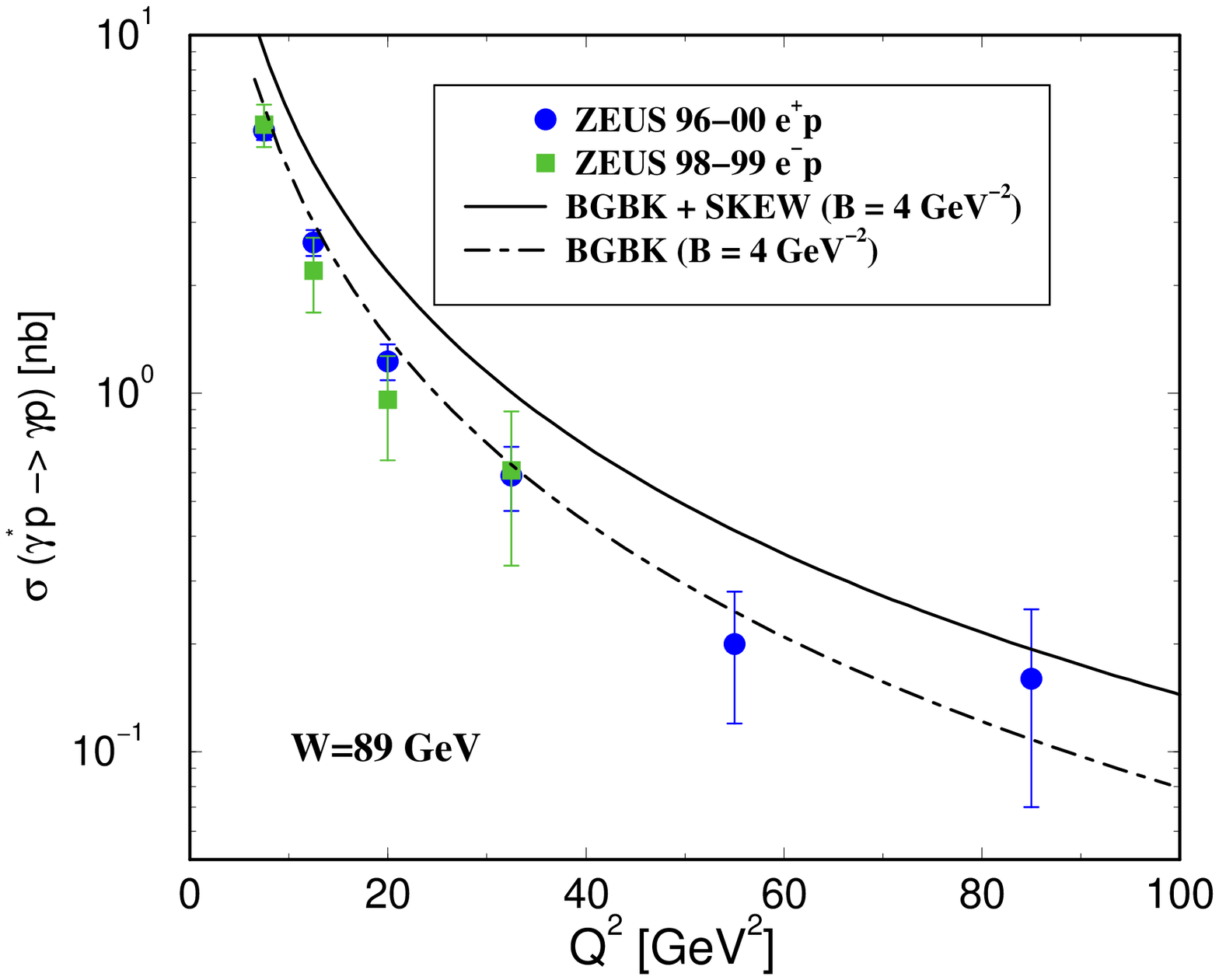,width=75mm}\\
(a) & (b)
\end{tabular}
 \caption{The results for the (a) saturation model with (full) 
 and without (dot-dashed) skewedness effect and (b) BGBK model, 
  with (dot-dashed) and without (full) skewedness effect
  for a fixed  $B=4$ GeV$^{-2}$.
  }
\label{fig:3} 
\end{figure*} 

In order to investigate whether a QCD evolution improves the description,
we show in Fig.~\ref{fig:2}-b the estimate using the BGBK 
dipole cross
section, Eq. (\ref{bgkdip}) as a function of $Q^2$ using fixed slope
values. There is an effect in the overall
normalization and a slower decrease at large $Q^2$ in contrast with the
model without QCD evolution reproducing well the ZEUS measurement for all 
$Q^2$. This suggests that DGLAP evolution starts to
be important for the large $Q^2$ points measured by ZEUS. 
A comparison of the different $Q^2$ behavior independently of the
normalisation question is presented at the end of this section.

Furthermore, we are motivated to investigate the importance of the
skewedness effects in the DVCS process using the previous results. Here,
we follow the approximation proposed in Ref.~\cite{Shuvaev:1999ce}, where
the ratio of off-forward to forward parton distributions are obtained
relying on simple arguments.
 The behavior of those ratios are given explicitly by \cite{Shuvaev:1999ce},
\begin{eqnarray}
  R_{q,g}\,(Q^2)=\frac{2^{2\lambda + 3}}
   {\sqrt{\pi}}\,\frac{\Gamma\,\left(\lambda+ 
    \frac{5}{2}\right)}{\Gamma \,\left(\lambda+3+p \right)}\,,
 \label{skew}
\end{eqnarray}
where $p=0$ for quarks and $p=1$ for gluons, and where $\lambda$ is
the exponent of the $x^{-\lambda}$ behavior of the input diagonal
parton distribution. It should be noticed that the skewed effect is
much larger for singlet quarks than gluons. In the following, it will
be assumed that the DVCS cross section is lead by a two gluon
exchange. In our further computations, we use $\lambda=\lambda(Q^2)$
as discussed in the previous section and the skewedness effect is
given by multiplying the total cross section by the factor
$R_g^2(Q^2)$. Once the effective power increases as a function of
$Q^2$, the skewedness effects could enhance the cross section by a
factor two if values of $\lambda_{eff}\simeq 0.4$ are reached at larger 
virtualities. In Fig.~\ref{fig:3}-a  we show the
result using the saturation model (4-flavor) and the skewedness
correction, Eq.~(\ref{skew}). The
same analysis is shown for the BGBK model in Fig.~\ref{fig:3}-b. The
main effect is to increase the overall normalization of the cross
section by about 40\% and only slightly modify the large $Q^2$ behaviour.
Again, this will be shown more clearly and independently of the
normalisation at the end of this section.

\begin{figure}[t]
\psfig{file=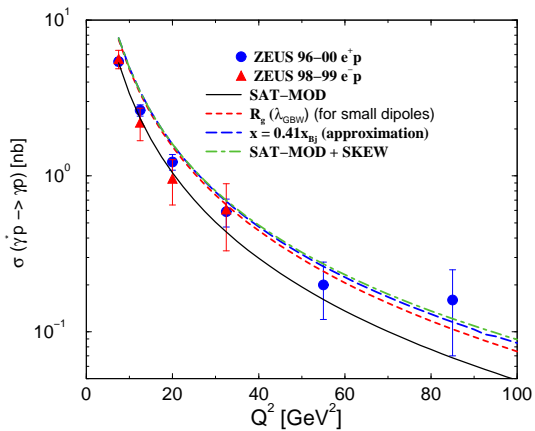,width=70mm}
\caption{Comparison among different approximations for the skewedness
  correction (see text).}
\label{compskews} 
\end{figure} 

For completeness, we have investigated two additional versions of
the implementation of the skewedness correction factor. They are shown in
Fig.~\ref{compskews} for fixed $B=4$ GeV$^{-2}$. First, we have
imposed skewedness correction only for small dipoles by introducing
$R_g(\lambda_{GBW}=0.277)$ in the exponent of
Eq.~(\ref{gbwdip}) (dotted line). This is to prevent correction to the
nonperturbative (large dipoles) piece of the dipole cross section. 
Further, we also test the rough
approximation $\tilde{x}=0.41\,x_{\mathrm{Bj}}$ (dashed line), which comes
from a simplified hypothesis 
$\sigma_{dip}\sim R_g(\lambda)\,\left(x_{\mathrm{Bj}}\right)^{-\lambda}$. 
The conclusion
is that these two different implementation of the skewedness correction
do not make sensible changes w.r.t.\,the first skewedness correction
neither in normalisation nor in $Q^2$ dependence for the presently covered 
kinematic range and precision of the measurement. 

At this stage, some comments are probably needed. 
The estimate for skewedness taken into
account above is an approximation as currently we have no accurate
theoretical arguments how to compute it from first principles within
the color dipole formalism. A consistent approach would be to compute
the scattering  amplitude in the non-forward case (the non-forward
photon wave function has been recently obtained in Ref.~\cite{Bartels:2003yj}).
In this case, the dipole cross section, $\sigma_{dip}(x_1,x_2,\rr,
\vec{\Delta})$,  depends on the light cone momenta
$x_1$ and $x_2$ carried by the exchanged gluons, respectively, and on
the total transverse momentum transfer $\vec{\Delta}$ (
additional information about the behavior on $\vec{\Delta}$ is needed for the
QCD Pomeron and proton impact factor). 
The forward dipole cross section is recovered at $x_1=x_2$ and
$\vec{\Delta}=0$. In the future, an experimental constraint for the
nonforward dipole cross section should be feasible with increasing
statistics on DVCS and exclusive (diffractive) vector meson
production.   
\\

\begin{figure*}[t]
\begin{tabular}{cc}
\psfig{file=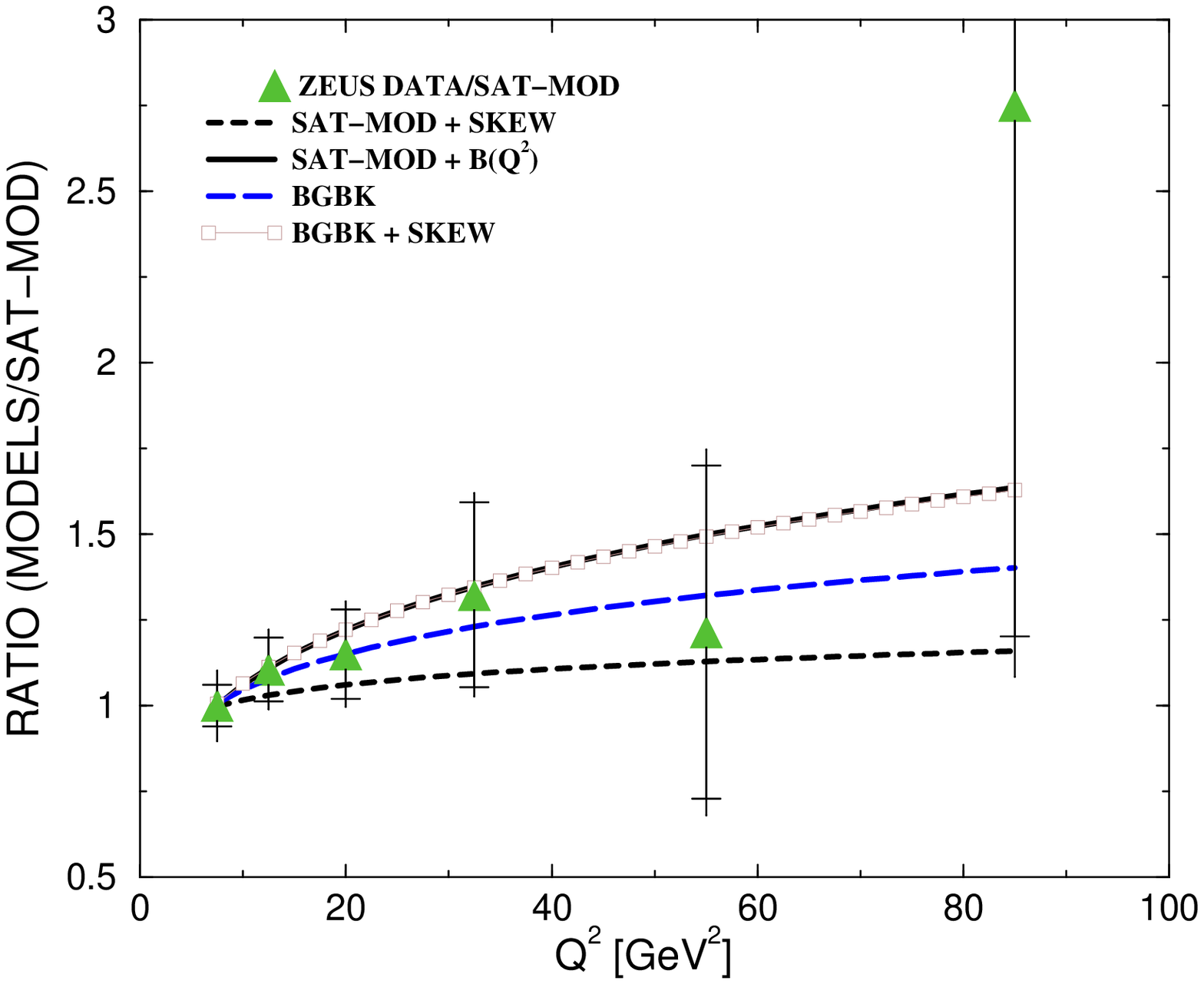,width=75mm} & \psfig{file=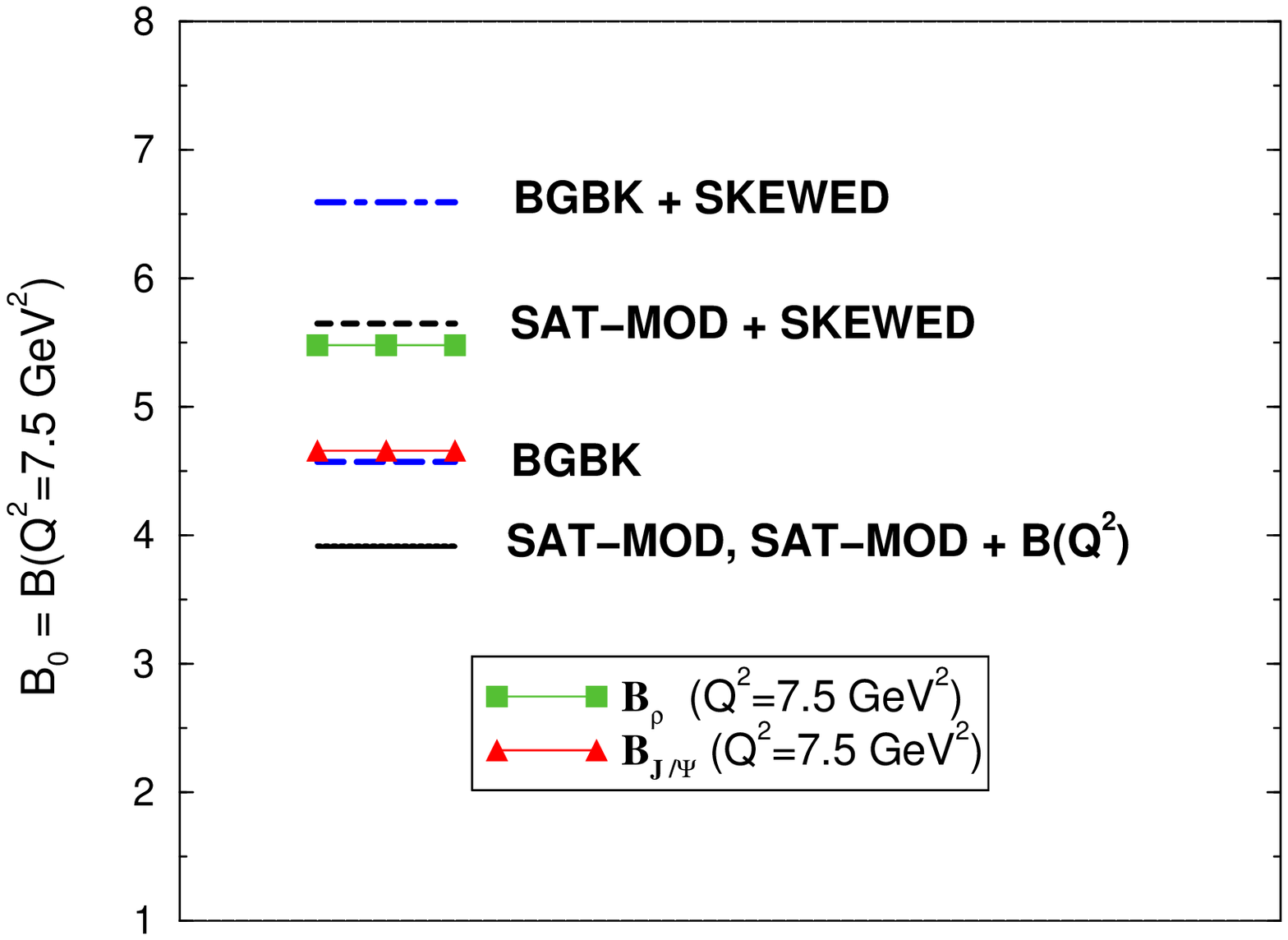,width=75mm}\\
a) & (b)
\end{tabular}
\caption{The ratio (a) MODELS/SAT-MOD as a function of $Q^2$ (models
  normalized to ZEUS data at $Q^2=7.5$ GeV$^{2}$), and (b) slope
  values of the models  at  $Q^2=7.5$ GeV$^2$ and slopes of heavy and
  light vector mesons.}
\label{fig:4} 
\end{figure*} 

To close this section, as the slope parameter $B$ has never been
measured for DVCS, we compare the different estimates 
presented in a systematic way separately for the effect on the $Q^2$
dependence and the effect on the overall normalisation.
To compare the $Q^2$ dependences, we normalize all models to describe
the ZEUS data point at the lowest $Q^2$ value, i.e.\, $Q^2=7.5$ GeV$^2$. 
Further, we plot the ratio of each model to
our baseline model SAT-MOD as a function of $Q^2$. 
Such a procedure allows a $Q^2$ dependence
comparison independently of the normalization effect. These ratios are
shown in Fig.~\ref{fig:4}-a, where the points (triangles-up) are the
ratio of the ZEUS data to SAT-MOD including the error bars for the
statistical (inner) and sum in quadrature of statistical and
systematic (outer) uncertainties. 

On the other hand, to compare the effect on the
normalisation we show the slope value needed to describe the lowest
$Q^2$ value of the ZEUS data points $B_0=B(Q^2=7.5$ GeV$^2$). 
They are shown in Fig.~\ref{fig:4}-b. 
For completeness, we also present the measured slope values for
vector meson production at that virtuality, both for $\rho^0$ 
and $J/\Psi$ mesons as indications of typical
values for respectively light and heavy mesons using the simple
parametrisation: 
\begin{eqnarray}
B = 0.60\,\left(\frac{14}{\left(Q^2+M_V^2\right)^{0.26}} + 1 \right)\,,
\end{eqnarray}
where $M_V$ is the meson mass.

From these comparisons, we conclude that several models can account
for the measured $Q^2$ dependence (SAT-MOD+B(Q$^2$), SAT-MOD+SKEW and BGBK, as
well as combination of several of those effects) which are not distinguishable
with the present experimental precision.
The difference between the models is much more pronounced in the
prediction of the cross section value, or in other terms, in the $B$
value needed to describe the integrated cross section over the 
available $Q^2$ range.
If the change in normalisation is small for the inclusion of a $Q^2$ 
dependence in $B$, the effect is of the order of 12\% for BGBK 
with respect to the basic SAT-MOD and of
40\% for the skewedness effect (SKEW) and still larger when the different
effect are combined (60\% for BGBK+SKEW).   

In summary, these issues show clearly the importance of a measurement
of the $t$ slope parameter $B$.

\section{Summary}
It has been shown that the DVCS cross section at HERA can
be described by the simple picture rendered by the color dipoles
formalism. In particular, the saturation model does an excellent job in
the current experimental kinematic domain. To achieve a good description
of the data up the the highest $Q^2$, the
original saturation model can be supplemented by QCD evolution, an 
additional dependence of B on $Q^2$ and skewedness effects. 
These effects modify in a sensitive way the absolute cross section
(10-60\%). Measurement of the $t$ slope parameter $B$ would already allow
to discriminate among the different theoretical predictions with
an amount of data comparable to the present ZEUS measurement.

\begin{acknowledgments}
 M.V.T.M. thanks the CERN Theory Division, where part of this work was
performed, for the hospitality and financial support.  
The work of L. Favart is supported by the FNRS of Belgium 
(convention IISN 4.4502.01) and  M. Machado was partially financed 
by the Brazilian funding agency CNPq.

\end{acknowledgments}

\end{document}